\documentclass{aa} 
\usepackage{natbib}
 \usepackage{color}
     \usepackage{xcolor}
\usepackage{ gensymb }
\bibpunct{(}{)}{;}{a}{}{,}
\usepackage{placeins} 
\usepackage{txfonts}
\usepackage{subcaption}
\usepackage{lscape} 
\usepackage{graphicx}
\usepackage{ wasysym }
\usepackage{amsmath}

\begin{document}

   \title{Energy deposition in planetary and exoplanetary atmospheres induced by cosmic rays}
   
   \subtitle{}

   \author{J. Polman
   \and
    I. Leya
    }

   \institute{Division of Space Research and Planetary Sciences, Physics Institute, University of Bern, Gesellschaftsstrasse 6, 3012 Bern, Switzerland}
             
   \date{Received 11 June 2026; accepted 3 September 2026}

\abstract
{Cosmic rays can significantly alter the abundances of certain species in the upper layers of planetary atmospheres, especially in terms of their biosignatures. To fully understand the extent of this effect, it is essential to accurately model the interactions of cosmic rays with planetary magnetic fields. We used the {\tt CosmicTransmutation} code to study the effect of a planetary magnetic field on the flux of galactic cosmic rays and stellar energetic particles. We found that both particle sources are significantly affected by magnetic fields, even though the effects are different due to the varying energy ranges that characterize each source. The stellar energetic particle energy flux is significantly higher for an Earth-like planet with no magnetic field, but with a magnetic field of 30\,$\mu$T or higher, the energy flux of the two sources becomes comparable. We find that the atmospheric pressures the cosmic rays reach are significantly lower than those found in previous studies using simpler models and that the effect of the atmospheric composition on this outcome is small. We found similar results using the model for the exoplanet K2-18b, even though the radius is significantly larger than Earth's. Because the two cosmic ray sources cover different energy ranges and are affected by the magnetic field in different ways as a result, both sources should be considered. Future studies should focus on combining an accurate modelling procedure of the interaction between cosmic rays and planetary magnetic fields with atmospheric chemistry models. This will require general circulation models to capture the latitudinal and longitudinal dependencies in full.}
\maketitle
\nolinenumbers

\section{Introduction}
In recent decades, more than 6300 exoplanets have been discovered, with an additional $\sim$8000 candidates awaiting confirmation. Among the currently known exoplanets, approximately 50-60 are potentially habitable; namely, they orbit within the habitable zone of their host stars where liquid water could exist. Since its launch in December 2021, the James Webb Space Telescope (\textit{JWST}, \citealp{Gardner+2023}) has enabled, for the first time, detailed spectroscopic characterizations of exoplanetary atmospheres, including those of Earth-like planets (e.g. \citealp{Damiano+2024,Allen+2026}). Future space missions such as PLAnetary Transits and Oscillations of stars (\textit{PLATO}, \citealp{Rauer+2025}), Atmospheric Remote-sensing Infrared Exoplanet Large-survey (\textit{Ariel}, \citealp{Tinetti+2016}), Habitable Worlds Observatory (\textit{HWO}, \citealp{Feinberg+2026}), and Large Interferometer for Exoplanets (\textit{LIFE}, \citealp{Quanz+2022}), together with next-generation ground based observations such as the Extremely Large Telescope (ELT, \citealp{Padovani+2023}), are expected to further advance the characterization of exoplanetary atmospheres with unprecedented sensitivity and spectral resolution.

One of the primary scientific objectives of these observational efforts is the detection and interpretation of biosignatures, which are gases or other planetary features that may indicate the presence of biological activity, individually or through a specific combination. Examples and additional references can be found in, e.g. \citet{Barthetal_2024}. However, robust interpretation of potential biosignatures requires a comprehensive understanding of abiotic processes capable of producing similar observational signatures \citep{Sousa_Silva_2020,Rodgers-Leeetal_2021, Seager+2025}.

Such processes may lead to false-positive detection by generating molecules or compounds commonly associated with life, or false-negative detections by masking or destroying genuine biosignatures. In addition to false-positive and false-negative signals, there is also the question of antibiosignatures, which are signals that usually suggest that the planet is uninhabited \citep{Schwietermanetal_2019}. Consequently, a reliable assessment of atmospheric spectra from potentially habitable exoplanets demands that all relevant non-biological mechanisms capable of producing, modifying, or erasing biosignatures be systematically investigated and quantitatively constrained.

Several studies have already investigated the influence of cosmic ray irradiation on exoplanetary atmospheres. For example, \citet{RimmerHelling_2013} studied the effect of cosmic rays on the ionization of atmospheres of brown dwarfs and giant gas planets. In their work, particle transport was treated by solving the Boltzmann equation; however, nuclear interaction, including the production of secondary particles, were not taken into account. Instead, the cosmic ray flux was assumed to decrease exponentially with atmospheric column density. In a subsequent study, \citet{Griessmeier+2016} found that the impact of galactic cosmic rays on atmospheric chemistry strongly depends on the presence and strength of planetary magnetic fields, which can severely reduce the incident cosmic ray flux and, consequently, the resulting chemical effects in the atmosphere. This occurs already for magnetic fields significantly weaker than that of Earth, which has a strength of $\sim$31\,$\mu$T \citep{Macmillan+2010}, although it can vary from 23\,$\mu$T to 70\,$\mu$T, especially depending on geographic latitude. Their results therefore emphasize the importance of consistently considering planetary magnetic fields when modelling the interaction of energetic particles with exoplanetary atmospheres.

Studies by \citet{Mesquita+2021} and \citet{Rodgers-Leeetal_2021} investigated the effects of stellar and galactic cosmic rays on exoplanetary atmospheres. However, both studies focussed primarily on the well-studied M-dwarf GJ~436 and on the propagation of energetic particles within the astrosphere. In an earlier and more general study, \citet{Mesquita+2020} considered only stellar energetic particles and did not include particle transport within or interactions with the exoplanetary atmosphere itself. Subsequently, \citet{Mesquita+2021} and \citet{Rodgers-Leeetal_2021} examined the relative importance of stellar and galactic cosmic rays as a function of stellar rotation rate. While, both studies successfully characterized the energetic particle environments at the top of the planetary atmosphere, neither included detailed transport calculations of energetic particles within the atmosphere. Consequently, the resulting atmospheric ionization, secondary particle production, and chemical effects could not be quantified.

Recent studies have focussed on the role of stellar and galactic energetic particles in shaping the chemistry of exoplanetary atmospheres. In particular, \citet{Barth+2021} demonstrated that stellar energetic particles and X-ray and UV (XUV) radiation strongly affect the ionization state and chemical composition of an exoplanetary atmosphere. Their results indicate that solar energetic particles (SEPs) enhance the abundance of hydrocarbons and other organic molecules, including formaldehyde (CH$_2$O) and acetylene (C$_2$H$_2$), which are both important key precursors in prebiotic chemistry pathways leading to amino acids, such as glycine. Previously, \citet{Seguraetal_2010} showed that interactions between stellar particles and oxygen-rich atmospheres can significantly deplete the ozone layer, illustrating the importance of including energetic particle-driven processes in realistic atmospheric chemistry networks.

In parallel, \citet{Rodgers-Leeetal_2021} discussed the potential relevance of cosmic rays for prebiotic chemistry and the origin of life. Cosmic ray-induced ionization can drive the formation of reactive species such as NH$_4^+$, H$_3^+$, and H$_3$O$^+$, thereby influencing key chemical pathways in planetary atmospheres. At the same time, energetic particle interactions could also produce species that mimic biosignatures; for example, through NO$_x$-driven chemistry. This further highlights the risk of false-positive biosignature interpretations.

Comprehensive modelling efforts, such as the {\tt MOVES} framework developed by \citet{Barth+2021}, have combined stellar XUV irradiation, cosmic rays, and stellar energetic particles to study their collective impact on the atmosphere of the hot Jupiter HD~189733b. This approach couples radiative transfer, energetic particle transport, and a detailed chemical network in a self-consistent manner. However, it remains limited to a specific planetary case and does not include the effects of a planetary magnetic field, which can significantly alter the penetration depth and transport of energetic particles and thus the resulting atmospheric chemistry.

The present project adopts a more comprehensive approach to investigating the effects of both stellar and galactic cosmic rays on exoplanetary atmospheres. Using the state-of-the-art Monte Carlo toolkit {\tt Geant4}, we computed the energy deposition profiles in planetary atmospheres as a function of orbital distance, chemical composition of the atmosphere, and planetary magnetic field. The study is restricted to the calculation of atmospheric energy deposition and does not address the subsequent physical and chemical consequences of this energy input. In particular, we do not couple the resulting energy deposition profiles to ionization calculations or atmospheric reaction network models. Such extensions are deferred to future work.

\section{Methods}
\subsection{The computer-code system}
We used the {\tt Geant4}-based model {\tt CosmicTransmutation} \citep{Hirtz2019,Hirtz+2022} to compute the differential particle spectra of galactic cosmic rays (GCRs) and stellar energetic particles (SEPs) as a function of altitude for planetary atmospheres with varying chemical compositions, magnetic field strengths, and planetary environments. A key advantage of {\tt CosmicTransmutation} is that, instead of relying on effective cut-off rigidities \citep{Smart+2000}, it explicitly accounts for the focussing and deflection of cosmic rays in planetary magnetic fields. This enables a more accurate determination of the fraction of particles reaching the upper part of the atmosphere as a function of latitude \citep{Hirtz+2022}. The computer code is designed to generate all the input information required for {\tt Geant4} simulations and to help the user extract the results from the sometimes very large output files \citep{Hirtz2019}.

The GEometry ANd Tracking Monte Carlo toolkit ({\tt Geant4}) is a general-purpose particle transport model originally developed for high-energy physics applications \citep{Agostinellietal_2003}. It provides a comprehensive set of physics models covering all physical processes important for nuclear interactions and transport over a wide energy range (i.e. from a few meV for neutrons up to TeV for all particles). In addition, {\tt Geant4} is a well-developed open source program that allows the user to define the geometry, the particle sources, and other relevant physical aspects in a very flexible way.

For the present study, we adopted the physics list FTFP$\_$INCLXX$\_$HP. At energies greater than 15 GeV for hadrons and greater than 2.9\,GeV\,nucleon$^{-1}$ for ions, we treated interactions using the Fritiof string model (FTF) in combination with the Preco precompound model \citep{NilssonAlmqvist1987, Andersson1987, Uzhinsky2011}. At intermediate energies we used the Li\`{e}ge Intranuclear Cascade Model (INCL) for the intranuclear cascade \citep{Boudard2013, Kaitaniemi2011}, in which nuclear reactions are represented as sequences of individual particle-particle collisions \citep{Serber_1947}.

The applicable energy range depends on the species of the particles; in {\tt Geant4.10.5}, INCL treats nucleons in the range of 1 MeV to 20 GeV, while for light ions, including alpha particles, it is applied between 0 and 3\,GeV\,nucleon$^{-1}$. Within the applied physics list FTFP$\_$INCLXX$\_$HP, INCL is applied only to neutrons down to 20\,MeV. At lower energies, neutron interactions are handled by the High Precision (HP) Neutron Model, which is based on evaluated nuclear data libraries and tabulated cross sections down to $10^{-12}$\,eV. The chosen physics list provides a consistent overlap between the individual models across energy regimes, ensuring smooth transitions and stable particle transport over the full energy range considered. More information can be found in \citet{Leyaetal_2021}.

The GCR and SEP spectra we adopted as input to the simulations are described in Appendix A. Examples of the corresponding spectra at the top of the atmosphere of an Earth-like planet are shown in Fig. \ref{Fig:Inputspectrum}. In the present study, we used the resulting  atmospheric particle fluxes and cascades solely to determine the spatial distribution of the deposited energy as a function of latitude and altitude. A detailed investigation of the associated atmospheric reaction processes, including ionization and chemically induced reaction networks, is beyond the scope of this work and will be addressed in a next step. We assumed a centred dipolar magnetic field, with the magnetic axis aligned with the planetary rotation axis. Under this assumption, the magnetic environment can be fully characterized by the surface dipole strength.

\subsection{Case 1: Earth-like planet}
We first considered an Earth-like planet to investigate the influence  of the magnetic field strength on the GCR and SEP spectra as a function of latitude. The simulated planet has a radius of 6378\,km and is surrounded by a 70\,km thick atmosphere. At the surface, the atmospheric density is 1.225\,kg\,m$^{-3}$ and deceases exponentially with altitude. Assuming a scale-height of 7.6\,km gives a 10$^{-4}$ times lower density at the top of the atmosphere (70\,km). Extending the atmosphere to lower pressures does not significantly affect the results, as most of the energy  deposition occurs at substantially higher pressures.

We used an atmospheric composition of 76.04\% nitrogen, 23.36\% oxygen, 0.58\% argon, and 0.02\% carbon by mass. To test the sensitivity of the modelled results to atmospheric density and chemical composition, we additionally considered atmospheres with densities scaled by factors 0.5 and 2, as well as idealized end-member cases consisting of a mixture of pure hydrogen and helium and an atmosphere consisting of pure CO$_2$.

We adopted a solar modulation parameter of $M=550$\,MV as a representative long-term average value for GCR modulation by solar activity, which is only slightly different from the value of $M=660$\,MV proposed by \citet{Leyaetal_2021}. The solar modulation parameter indicates the average decrease in rigidity of GCRs travelling through the heliosphere. This parameter is given in MV (megavolts), because the energy loss of the particles is dependent on their charge. To explore the impact of solar modulation, we additionally considered values of 100\,MV and 1000\,MV, corresponding to low and high solar activity conditions, respectively.

\subsection{Case 2: K2-18b}
To investigate the impact of planetary magnetic fields on the GCR and SEP spectra incident on exoplanets, we considered K2-18b \citep{Montet+2015}. K2-18b is an interesting target, because of its presence in the habitable zone and the possible presence of biosignatures in its atmosphere \citep{Madhusudhan+2023,Wogan+2024,Madhusudhan+2025,Stevenson+2025}. The planet has a radius of 2.61\,R$_\oplus$ \citep{Benneke+2017,Benneke+2019}, which is adopted directly in our model. We assumed an atmospheric composition of 42\% carbon, 41\% hydrogen, and 17\% oxygen by mass, based on the atmospheric models of \citet{Schmidt+2025}, which reproduce the \textit{JWST} observations of \citet{Madhusudhan+2023} with a 100\,$\times$ solar metallicity atmosphere, a C/O ratio of 3.25, and nitrogen depletion. To determine the atmospheric scale height, we combined the planetary radius with a planetary mass of 8.63\,M$_\oplus$ \citep{Cloutier+2017,Cloutier+2019}. Furthermore, we adopted a mean molecular weight of 6.5\,amu, consistent with an atmosphere dominated by CH$_4$, O$_2$, and H$_2$ mixtures and also within the constraints of $\mu_{\rm atm}=2.46-7.64$\,amu derived by \citet{Schmidt+2025}.

Assuming an atmospheric temperature of 340\,K \citep{Madhusudhan+2025, Schmidt+2025} and hydrostatic equilibrium, we obtained a scale height of $H \approx$ 35\,km. In our simulations, we adopted a reference atmospheric density of 1.0\,kg\,m$^{-3}$ at the lower boundary (bottom of the atmosphere). This value does not imply a physical planetary surface at this pressure level; it only defines the lower limit of the simulated particle transport. We defined the top of the atmosphere (upper boundary) at a density reduced by a factor of 10$^{-4}$ relative to this reference value, which corresponds to a total atmospheric thickness of $\sim$325\,km. This does not affect our results, since we find that no cosmic ray particles reach our imposed bottom of the atmosphere. 

K2-18 is a M2.5 dwarf with a mass of 0.495\,M$_\odot$ \citep{Montet+2015,Cloutier+2019} and an estimated age of 2.9--3.1\,Gyr old \citet{Sairam+2025}. To estimate the SEP flux, we adopted some scaling relationships. We started with the results of \citet{Mesquita+2021} for star GJ~436, which are  based on the stellar wind model of \citet{Mesquita+2020}. GJ~436 is a close analogue to K2-18, also being a M2.5 dwarf with a mass of 0.492\,M$_\odot$ \citep{Torres2007}, although it is older (8.9$^{+2.3}_{-2.1}$\,Gyr). We adopted case A of \citet{Mesquita+2021}, with a stellar wind terminal velocity of v$_\infty$=1250\,km/s and a stellar mass-loss rate of $\dot{M}=1.2\cdot10^{-15}$\,M$_\odot$\,yr$^{-1}$. The stellar wind kinetic power is given by $P_{SW}=\dot{M}v^2_\infty$. Assuming that the fraction of stellar wind power converted to SEPs is similar for K2-18 and the Sun, we can scale the SEP flux using solar reference values for the solar wind terminal velocity and the mass loss rate. Adopting v$_\infty=430$\,km\,s$^{-1}$ \citep{LeChat+2012} and $\dot{M}=2.1\cdot10^{-14}$\,M$_\odot$\,yr$^{-1}$ for the Sun yields an SEP flux scaling factor of 0.4829 relative to the solar value. We therefore adopted this scaling for K2-18. The SEP spectrum was assumed to be identical to the solar spectrum described in Appendix A, with the flux received by K2-18b further rescaled according to the orbital distance of 0.1591\,au \citep{Benneke+2019}. For the GCR spectrum, we also adopted the Earth-like spectral shape, but use a solar modulation potential of $M=600$\,MV,  estimated from the results  of \citet{Mesquita+2021} for GJ~436 and adjusted for the orbital distance of K2-18b. This value of the solar modulation parameter only differs slightly from that of Earth and is thus unlikely to have a significant effect on the results. 

No direct measurements of the magnetic field of K2-18b are currently available and only tentative indirect constraints exist for a limited number of exoplanets \citep{Turner+2021,Turner+2023}. We therefore considered three magnetic field scenarios for K2-18b. First, we assumed that the planet has no magnetic field. Second, we estimated the intrinsic magnetic field of K2-18b using the scaling law of \citet{Sano1993}, $B^2\propto\rho\Omega^2R_c$, where $\rho$ is the core density, $\Omega$ the rotation rate, and $R_c$ the core radius. This approach minimizes the number of free parameter compared to more complex dynamo models \citet{Christensen2010}.

Given the poorly constrained interior structure of K2-18b \citep{Madhusudhan+2020}, we assumed a core radius of 1\,R$_\oplus$ and an Earth-like core density, acknowledging that these values are uncertain. Based on the age and planetary parameters of K2-18, the planet is likely tidally locked (e.g. \citealp{Griessmeier2005}), implying a rotation period equal to the stellar rotation period of 38.6\,days \citep{Cloutier+2017}. Calibrating the scaling law to Earth with $B_\oplus=30$\,$\mu$T, we obtained an estimated magnetic field strength for K2-18b of $B_{\rm K2-18b}=1.7$\,$\mu$T.

Since it is unclear how accurate the scaling law and our assumptions are, we additionally considered as a third scenario representing an upper-limit on the planetary magnetic field-strength based on the results of \citet{Kilmetis+2024}. We adopted a surface magnetic field strength of 300\,$\mu$T, corresponding to the maximum value predicted for a planet with a similar mass as K2-18b, a similar envelope fraction of 10\%, a comparable a semi-major axis of 0.2\,au, and a similar age of 1.5\,Gyr (Fig. 5 of \citealp{Kilmetis+2024}). Although these parameters do not exactly match those of K2-18b and assume that the planet has a substantial metallic core, this scenario provides a plausible upper limit on the magnetic shielding effect of the planet on incident cosmic rays. 

\subsection{Simplifications and limitations for the model} \label{Sec:Simplifications}
We assumed that the incident GCR and SEP fluxes are spatially homogeneous and do not distinguish between day- and nightsides of the planet. While this approximation is reasonable for most freely orbiting planets, it is likely wrong for tidally locked planets. In such systems, the dayside is expected to be more strongly exposed to stellar energetic particles but is mostly shielded from galactic particles. In contrast, the nightside likely receives a larger relative contribution from galactic cosmic rays. Consequently, significant longitudinal variations in the particle environment may arise that are not captured by the present model.

In addition, we assumed that the incoming stellar flux is isotropic. In reality, stellar energetic particles are guided by the stellar magnetic field and propagate along spiral magnetic flux tubes. The resulting particle distribution is therefore generally anisotropic and can vary substantially with time. This anisotropy may affect both, the local SEP flux and the efficiency with which planetary magnetic fields shield the atmosphere, as shielding depths on the arrival direction of the particles.

We also ignored the obliquity of the magnetic field. This does not directly affect our results, since we assumed that the incoming stellar flux is isotropic, but it would have an effect if the incoming stellar flux would be anisotropic. \citet{Presa+2024} showed that the obliquity of the magnetic field can have a significant effect on the atmospheric escape rate due to solar-wind particles for hot Jupiters within the Alfv\'en radius.

Finally, energetic stellar particle events, like coronal mass ejections (CME), are frequently associated  with enhanced X-ray and extreme ultraviolet (EUV) emission. In the present study, we considered only the energy deposition caused by energetic cosmic ray particles and do not include the effects of X-ray and EUV radiation. Consequently, the calculations may overestimate the importance of magnetic shielding for the total atmospheric energy budget, particularly during periods of intense stellar activity when radiative energy deposition may become comparable to or even exceed the contribution from energetic particles.

\section{Results for case 1: Earth-like planet}
\begin{figure}
  \resizebox{\hsize}{!}{\includegraphics{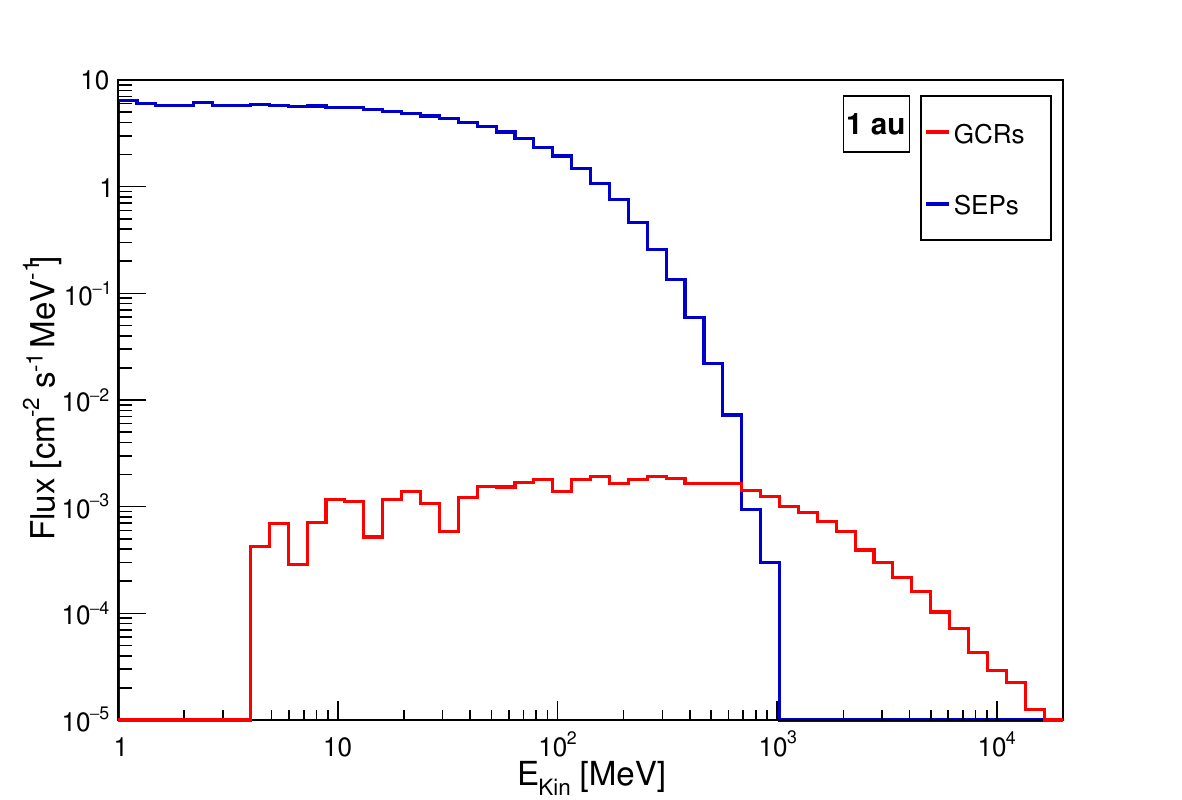}}
  \caption{Spectra of GCRs and SEPs at the top of the atmosphere for an Earth-like planet at 1\,au without magnetic field.}
  \label{Fig:Inputspectrum}
\end{figure}

Figure \ref{Fig:Inputspectrum} depicts the spectra of GCRs and SEPs incident on the atmosphere in the Earth-like planet case. The simulated spectra do not exactly reproduce the analytical distribution described in the Appendix because the particles are generated through  Monte Carlo sampling. To maintain computational tractability, the number of sampled particles was limited to 10$^4$ for the GCR simulations and 10$^5$ for the SEP simulations. Unless otherwise noted, these particle numbers were used in all simulations presented in this study. 

\subsection{Dependence on the magnetic field strength} \label{Sect:Mag_field_dep}
In the absence of a magnetic field, the spectra of cosmic rays entering the atmosphere are independent of latitude. For planets with magnetic fields, however, latitude becomes a key parameter due to the rigidity-dependent geomagnetic cut-off. In the following, we analyse the deposited energy as a function of altitude and latitude. The deposited energy is determined by taking the flux in cm$^{-2}$\,s$^{-1}$ and multiplying it by the energy of the particles, which gives the energy flux in MeV\,cm$^{-2}$\,s$^{-1}$. The flux is usually binned in energy when shown as a function of the particle energy, changing the unit of flux to cm$^{-2}$\,s$^{-1}$\,MeV$^{-1}$. 

\begin{figure*} 
\centering
   \includegraphics[width=17cm]{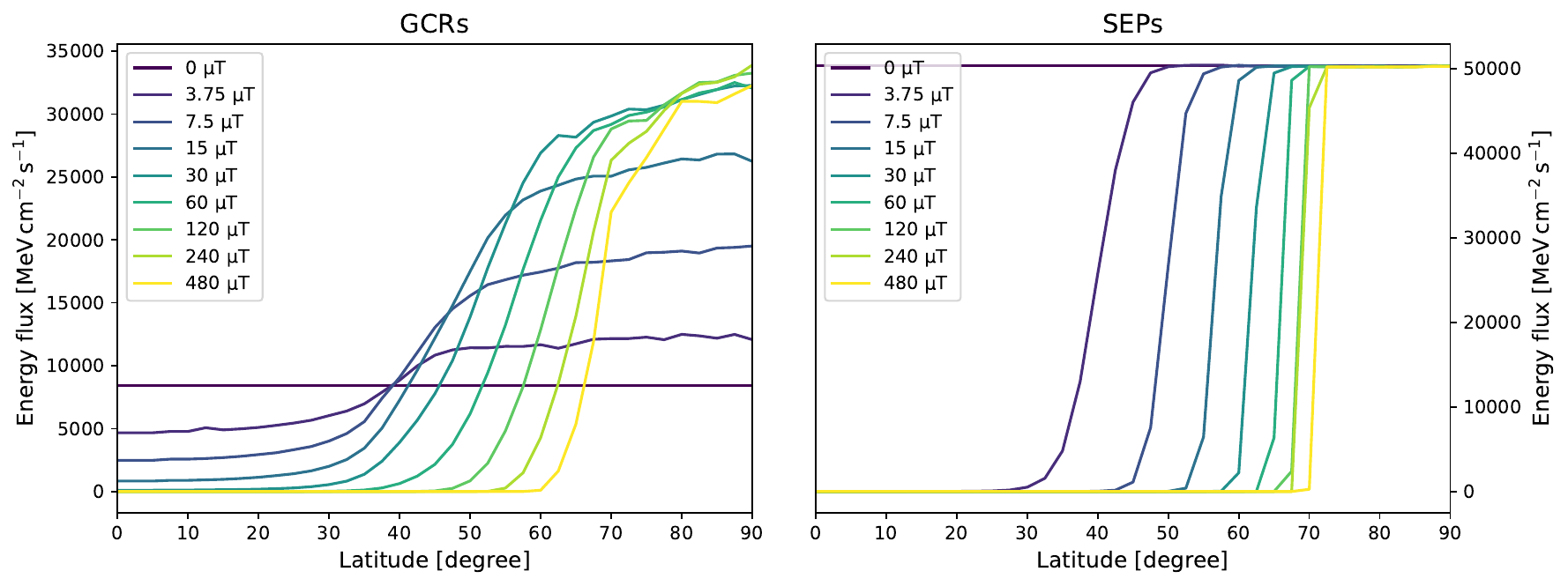}
     \caption{GCR (left) and SEP (right) energy flux at each latitude for varying magnetic field strengths for an Earth-like planet at 1\,au.}
     \label{Fig:LatitudeComparisonTwice}
\end{figure*}

The dependence of the deposited energy on latitude and magnetic field strength is shown in Fig. \ref{Fig:LatitudeComparisonTwice}, where the planet is sampled in steps of 2.5$\degree$. As discussed above, instead of analysing full energy spectra, we computed the energy flux at each latitude and subsequently averaged the data over the planetary surface.

A clear reduction in the energy flux at low latitudes is observed with increasing magnetic field strength. This effect is more pronounced for SEPs (right panel) than for GCRs (left panel). The reduction at low latitudes is primarily caused by the rigidity-dependent geomagnetic cut-off, which excludes an increasing fraction of low- and intermediate-energy particles and thereby reduces the number of particles reaching the upper atmosphere. Consequently, the deposited energy decreases in these regions.

However, a stronger magnetic dipole does not uniformly suppress particle access at all latitudes. Instead, it removes a large fraction of particles at low latitudes, while a subset of particles that are excluded from these regions still possess sufficient rigidity to penetrate the atmosphere at higher latitudes, i.e. their rigidities exceed the local polar cut-off. As a result, both the flux density and the deposited energy can increase in polar regions, as seen in the simulations. This behaviour reflects a reconfiguration of rigidity cut-offs rather than a simple global suppression of energetic particles, effectively shifting partially allowed trajectories towards higher latitudes. This trend is clearly visible in the left panel of Fig. \ref{Fig:LatitudeComparisonTwice}, which shows the energy deposited by GCRs as a function of latitude for different magnetic field strengths. The small fluctuation visible in the curves at high latitudes are caused by limited particle statistics (the total energy is carried by very few particles) and do not have any physical significance. 

For SEPs, the trend is somewhat different. As for GCRs, the deposited energy exhibits a strong dependence on latitude. For a magnetic field strength similar to that of Earth of 30\,$\mu$T, for example, SEPs are almost completely excluded below a latitude of 60$\degree$. However, in contrast to GCRs, there is little to no increase in the deposited energy at high latitudes with increasing magnetic field strength.

This different response of GCRs and SEPs to variations in the magnetic field strength is primarily due to differences in their spectral properties relative to the geomagnetic rigidity cut-offs. GCRs possess a broad, hard spectrum with a significant fraction of particles in the intermediate rigidity range where geomagnetic access is highly sensitive to both latitude and geomagnetic field strength. Consequently, changes in the dipole field modify the cut-off structure and redistribute particle access in phase space, which can lead to an increase of deposited energy at high latitudes despite the overall reduction in flux. 
 
The GCR spectrum does contain high energy particles, which is why an enhancement in the energy flux can be observed at higher latitudes when a magnetic field is included. For Earth this enhancement seems to reach a maximum for 30\,$\mu$T, with stronger magnetic fields only further limiting the latitudes where the enhancement occurs without increasing the enhancement itself. Because the GCR spectrum extends to sufficiently high rigidities, a significant fraction of particles excluded at lower latitudes can still penetrate the atmosphere in polar regions, resulting in the observed enhancement of the deposited energy. For the Earth-like case, this enhancement appears to reach a maximum for a magnetic field strength of approximately 30\,$\mu$T. Stronger magnetic fields do not lead to a further increase in the deposited energy, but instead progressively suppresses the GCR flux at low latitudes and restrict the enhancement to a narrow range of high latitudes. In contrast, the SEP spectrum is considerably steeper and is dominated by low-energy particles that are either already excluded at low and mid latitudes or efficiently transmitted at high latitudes under weak fields. As a result, increasing the magnetic field primarily suppresses SEP access rather than redistributing it, and no comparable enhancement at high latitudes is observed.

To be more quantitative, we examine the contribution of different particle energy ranges to the total energy flux in the absence of a magnetic field. For GCRs, less than 0.1\% of the total energy flux is carried  by particles with energies below 100\,MeV. Particles in the energy range from 100\,MeV to 1\,GeV contribute approximately $\sim$9\%, while the remaining 91\% originates from particles with energies higher than 1\,GeV. In contrast, the SEP energy flux is dominated by substantially lower-energy particles; approximately 31\% of the total energy flux is contributed by particles with energies below 100\,MeV, while the remaining 69\% originates from particles 100\,MeV and 1\,GeV. No SEP particles with energies above 1\,GeV are present in the adopted spectrum. These differences clearly explain why geomagnetic shielding affects SEPs more strongly than GCRs and why higher-latitude enhancement observed for GCRs is absent for SEPs.

\begin{figure}
  \resizebox{\hsize}{!}{\includegraphics{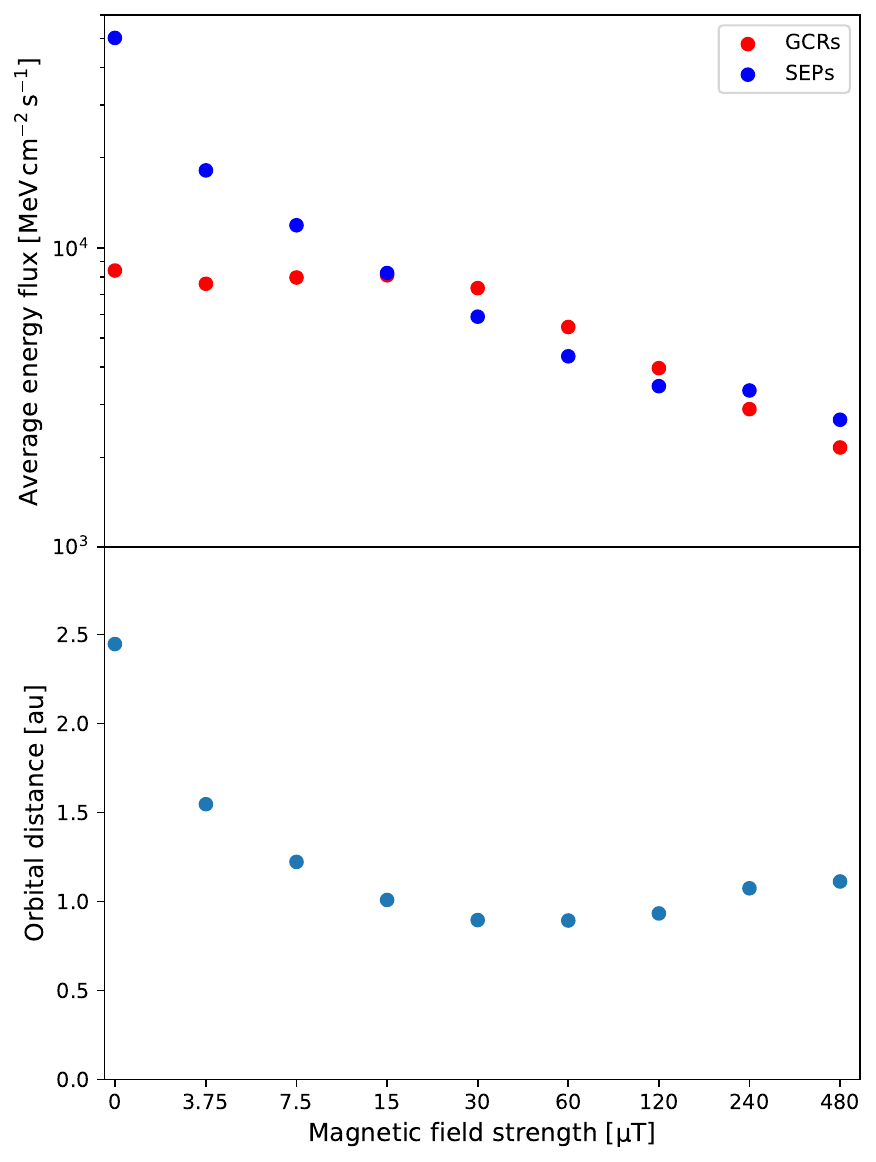}}
  \caption{Energy flux of GCRs and SEPs averaged over the surface of the planet (top) at a distance of 1\,au and the orbital distance at which the GCR and SEP energy flux would be equal (bottom) for varying magnetic field strengths for an Earth-like planet.}
  \label{Fig:EqualDistancePlot}
\end{figure}

Figure \ref{Fig:EqualDistancePlot} (upper panel) shows the energy fluxes of GCRs and SEPs averaged over the planetary surface, as a function of magnetic field strength for a planet at an orbital distance of 1 au. In this calculation, the GCR flux is assumed to be independent of orbital distance, neglecting variations in the solar modulation parameter, while the SEP flux is assumed to scale as $R^\alpha$, where $R$ is the orbital distance and $\alpha=-2$. Observational studies have reported values of $\alpha$ ranging from --1.4 to --3.7 \citep{Gardini+2008,Cao+2025}, depending on the energy range considered and whether the peak fluxes or fluences are analyzed. We adopt $\alpha=-2$ because \citet{Cao+2025} found that |$\alpha$| decreases with increasing particle energy, even within the relatively narrow energy range of 10.5--40\,MeV.

The figure demonstrates that SEPs dominate the globally averaged energy flux in the absence of a magnetic field. However, this changes rapidly as magnetic shielding becomes more effective. Up to a magnetic field strength of 30\,$\mu$T, the SEP energy flux decreases substantially, whereas  the GCR energy flux remains nearly unchanged. Consequently, the orbital distance at which GCR and SEP energy fluxes are equal decreases from $\sim$2.4\,au when no magnetic flux is present to $\sim$0.9\,au for a magnetic field strength of 30\,$\mu$T (lower panel of Fig. \ref{Fig:EqualDistancePlot}). For stronger magnetic fields, both GCR and SEP energy fluxes decrease. At this field strengths, the reduction in the GCR flux becomes more pronounced, causing the orbital distance where the two energy fluxes are equal to increase slightly, reaching $\sim$1.1\,au for a magnetic field strength of 480\,$\mu$T. The finite latitude resolution becomes increasingly important at larger magnetic field strengths because the SEP flux is confined to progressively narrower polar region.

These results indicate that neither GCRs nor SEPs dominate the atmospheric energy input at 1\,au in a general sense, particularly for planets with intrinsic magnetic fields. Instead, both particle populations contribute in a complementary way that depends on orbital distance and magnetic shielding. This is physically important because GCRs and SEPs differ not only in flux level but also in spectral shape, which determines the depth of energy deposition, the efficiency of secondary particle production, and ultimately the altitude range over which atmospheric ionisation occurs. As a result, variations in orbital distance or magnetic field strength do not simply rescale the total energy input, but can qualitatively modify where in the atmosphere energy is deposited and which physical processes are driven. With the adopted ($R^{-2}$) scaling of the SEP flux, SEPs are expected to dominate the high-energy particle environment of close-in planets, whereas GCRs become increasingly relevant at larger orbital distances, where they govern both the integrated energy input and the baseline ionisation structure of the atmosphere.

\subsection{Solar modulation parameter}
\begin{figure}
  \resizebox{\hsize}{!}{\includegraphics{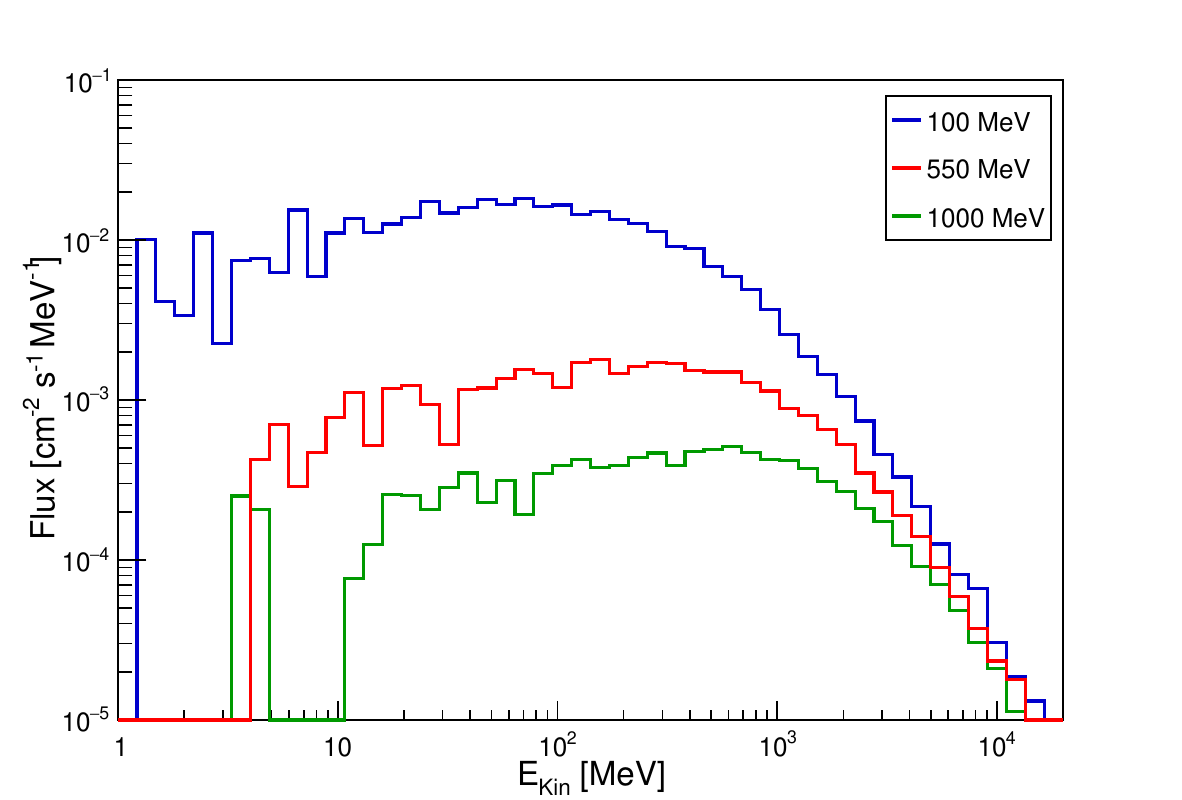}}
  \caption{GCR particle spectra for different values for the solar modulation potential ($M = 100$, 550, and 1000\,MV)}
  \label{Fig:SolarModulationComparison}
\end{figure}

Up to this point, we neglected variations in the solar modulation potential. This assumption was adopted both to simplify the analysis and because the dependence of the solar modulation potential is not always well constrained, particularly for an exoplanetary system. Even within the Solar System, the modulation potential exhibits substantial temporal variability associated with the solar activity cycle \citep{Usoskin+2011}. In addition, the solar modulation potential depends on orbital distance even for a steady-state stellar wind, with stronger modulation occurring close to the host star. Figure \ref{Fig:SolarModulationComparison} illustrates the resulting GCR spectra for three different values of the solar modulation potential: 100\,MV, 550\,MV (our reference case), and 1000\,MV. These values cover the range of possible modulation conditions and approximately cover the range observed over the solar cycles at Earth \citep{Usoskin+2011}.

As expected, the influence of solar modulation is strongest at low particle energies, although measurable differences persist up to higher energies. Relative to the reference case of 550\,MV, reducing the solar modulation potential to 100\,MV increases the globally averaged GCR flux by approximately 87\%, whereas increasing it to 1000\,MV decreases the energy flux by about 35\%. These results suggest that neglecting orbital-distances-dependent changes in the solar modulation potential does not quantitatively alter the conclusions drawn from the lower panel of Fig. \ref{Fig:EqualDistancePlot}. Because the SEP energy flux scales as $R^{-2}$ and the GCR flux changes only little with orbital distance ($\equiv$ solar modulation), even relatively large changes in the orbital distance ($\equiv$ solar modulation) translate into modest shifts in the equal-flux distance. Furthermore, the range of solar potentials considered here corresponds to a significantly broader range of conditions than those associated with the orbital distances explored in Fig. \ref{Fig:EqualDistancePlot}. Consequently, the resulting variations in the energy flux provide a very conservative estimate of the maximum impact that the orbital-distance dependency on the solar modulation potential might have on the results.

\subsection{Atmospheric deposition}
\begin{figure*} 
\centering
   \includegraphics[width=17cm]{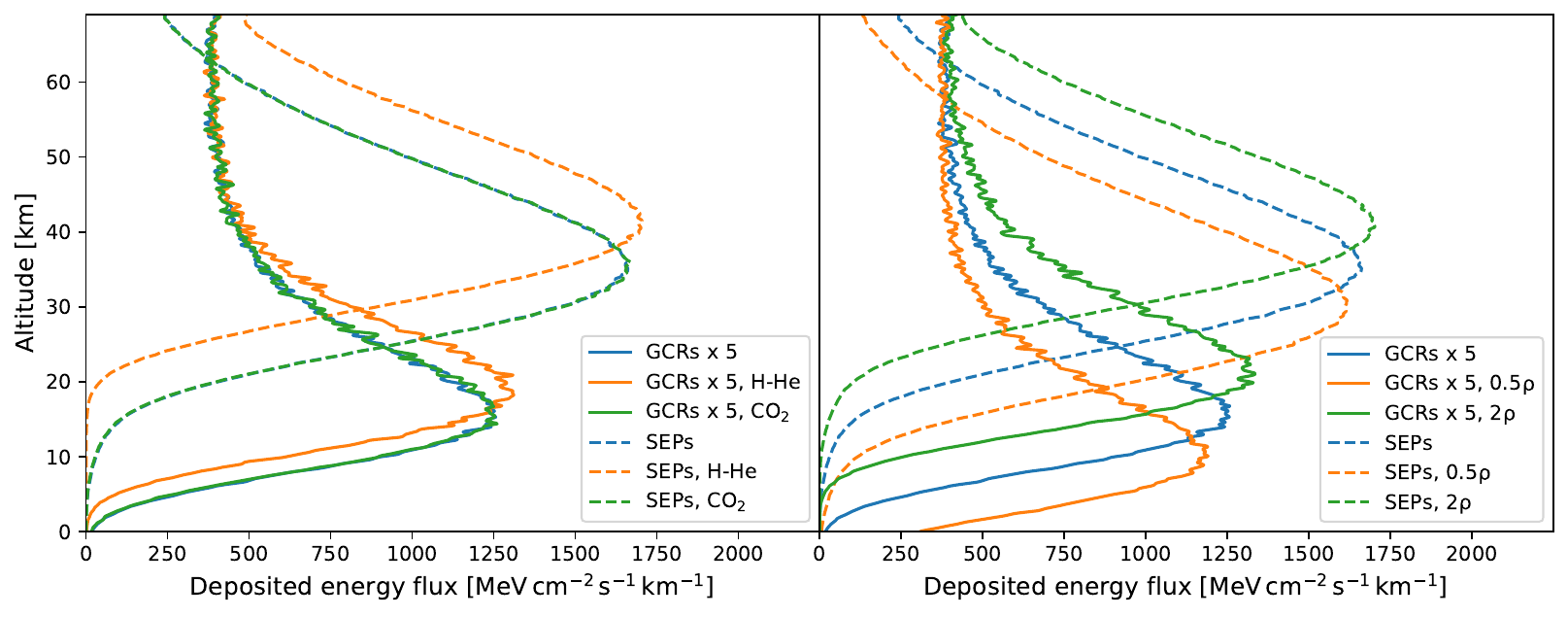}
     \caption{Atmospheric energy deposition by GCRs and SEPs for the standard model, compared to atmospheres consisting purely of H-He and CO$_2$ (left) and to atmospheres with half and twice the atmospheric density (right), for an Earth-like planet without a magnetic field at 1\,au. The curves have been smoothed, and the deposited energy for GCRs multiplied by 5 for improved visibility.}
     \label{Fig:AtmosphereCompositionComparison}
\end{figure*}

Figure \ref{Fig:AtmosphereCompositionComparison} shows the energy deposition profiles as a function of altitude for the standard Earth-like case without a magnetic field. To properly resolve the altitude dependence of the energy deposition, these simulations were performed with a larger number of particles, 10$^6$ for the GCR spectrum and 10$^7$ for the SEP spectrum. For the spectra that were used, these numbers are still smaller than the number of particles in\,m$^{-2}$\,s$^{-1}$. Because a significant fraction of the energy flux is carried by a small number of high energy particles, and because of the randomness inherent to the energy deposition process, the energy deposition profiles have been smoothed.

We considered the standard terrestrial atmospheric composition and density, as well as atmospheres composed entirely of hydrogen-helium (H-He) or CO$_2$ (left panel of Fig. \ref{Fig:K218bComparisonDeposition}). Although it is not physically realistic to assume identical density profiles for atmospheres with different compositions or to scale the atmospheric density uniformly at all altitudes, these simplified models provide a useful framework for isolating and comparing the effects of atmospheric composition and density.

In all cases, GCRs deposit their energy deeper in the atmosphere than SEPs, which is expected because GCRs possess significantly higher characteristic energies than SEPs. This finding is consistent with previous studies \citep{Barth+2021,Rodgers-Lee+2023}. For the standard atmosphere, the deposition of energy from the GCR (blue line) exceeds that of the SEP (orange line) at a pressure of $\sim$10$^{-1}$\,bar, corresponding to an altitude of $\sim$15\,km. This transition occurs at a lower pressure than reported by \citet{Rodgers-Lee+2023}. This difference is most likely due to the fact that those authors considered orbital distances between 0.01-0.20\,au. At such distances, the SEP flux is substantially enhanced and with higher SEP energy fluxes, there is more energy deposition deeper into the atmospheres and therefore to higher pressures. Under Earth-like conditions, only a small fraction of GCRs reach the planetary surface, while the corresponding fraction of SEPs is negligible.

Changing the atmospheric composition to a pure H-He mixture (green line for GCRs and red line for SEPs) shifts the energy deposition to higher altitudes. Because the vertical density profile is held fixed, the lower molecular weight, $m$, results in a larger number density, $n$, via $\rho = n \cdot m$. The mean free path of incoming cosmic rays is inverse proportional to the number density and the interaction cross section, $\sigma$, via $\lambda = (n \cdot \sigma)^{-1}$, leading to a reduced geometric interaction length in physical distance.

The energy loss of protons in matter is mainly dependent on the electron density of the medium. The terrestrial atmospheric constituents (N$_2$, O$_2$) contain approximately 14-16 electrons per molecule, while H$_2$ contains only two electrons per molecule. However, when normalized by molecular mass, one obtains $Z/A \sim$ 0.5 for N$_2$ and O$_2$, while $Z/A = 1$ for hydrogen. Consequently, H-He atmospheres have approximately twice the electron density per unit mass compared to N$_2$-O$_2$ atmospheres, leading to moderately enhanced ionisation energy losses per unit mass column. In addition, hydrogen has a relatively low mean excitation energy, which further increases the stopping power through its logarithmic contribution in the Bethe-Bloch formalism. However, this effect is secondary compared to the dependence on electron density and atmospheric column mass. Overall, these factors lead to enhanced proton energy deposition and a corresponding shift of the deposition profile towards higher altitudes in H-He atmospheres compared to the standard terrestrial atmosphere. This finding is in excellent agreement with previous studies \citep{Atrietal_2013,RimmerHelling_2013,Barth+2021,Rodgers-Lee+2023}. 

For secondary neutrons, the fractional energy loss per collision scales as 2$A$/($A$+1)$^2$, where $A$ is the target mass number. Consequently, the higher particle number density at a given mass density, combined with the lower average atomic mass of an H-He atmosphere, leads to more efficient energy deposition at higher altitudes  than in the terrestrial atmosphere, in agreement with the model results. In contrast, replacing the terrestrial atmospheric composition with pure CO$_2$ produces only minor changes in the energy deposition profile, with the corresponding curves being nearly indistinguishable from the standard case. This behaviour is consistent with the discussion presented above.

The effect of atmospheric density is straightforward to interpret. Increasing the density (at fixed chemical composition) causes a larger fraction of the cosmic-ray energy to be deposited at higher altitudes, whereas decreasing the density shifts the energy deposition to deeper atmosphere layers, allowing a significantly larger fraction of the GCR flux to reach the surface. Overall, the influence of atmospheric properties on the energy deposition profile remains relatively modest because the atmospheric density itself varies by approximately four orders of magnitude between the top and bottom of the atmosphere. Consequently, even substantial changes in the density profile result in only a shift of a few kilometres in the altitude of peak energy deposition.  

It should be noted that the energy deposition from cosmic rays is substantially lower than the the energy from other sources, such as XUV radiation. Cosmic rays are unique in that they penetrate deeper into the atmosphere than other energy sources and that individual particles have large energies.

\section{Results for case 2: K2-18b}
\begin{figure*} 
\centering
   \includegraphics[width=17cm]{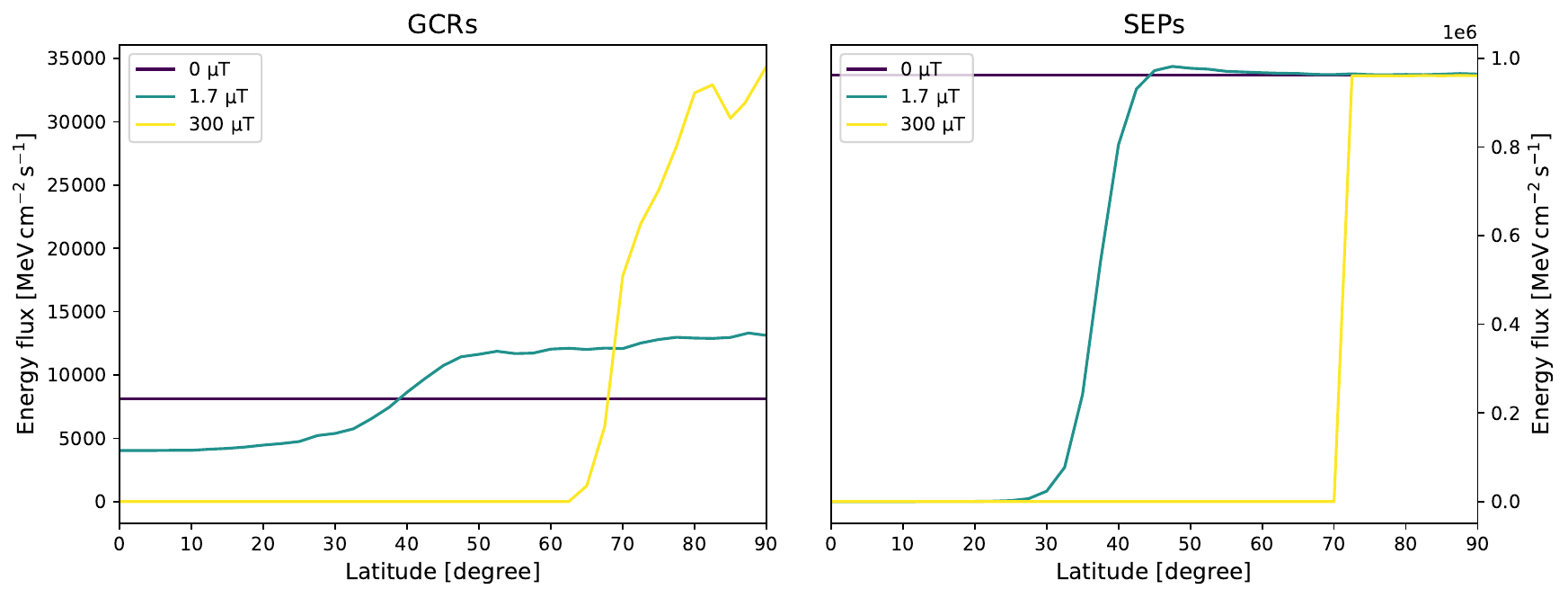}
     \caption{Latitudinal distribution of the GCR (left) and SEP (right) energy fluxes in the atmosphere of K2-18b for different magnetic field strengths.}
     \label{Fig:K218bLatitudeComparison}
\end{figure*}

Figure \ref{Fig:K218bLatitudeComparison} shows the energy flux of GCRs and SEPs entering the atmosphere of K2-18b for the different magnetic field strengths considered in this work. The fluctuations in the energy flux for a magnetic field strength of 300\,$\mu$T at high latitudes are caused by limited particle statistics, which is also shown in Fig. \ref{Fig:LatitudeComparisonTwice}. In general, the relative latitudinal distribution of the energy flux is similar to that found for the Earth-like case. However, the impact of a given magnetic field strength appears to be slightly stronger for K2-18b, likely due to its larger planetary radius. Larger planets have systemically higher rigidity cut-offs at the surface, leading to a stronger suppression of low-rigidity particles and an enhanced latitude dependence of atmospheric particle access. Test simulations varying the planetary radius support this interpretation. The results for magnetic field strengths of 1.7\,$\mu$T and 300\,$\mu$T for K2-18b correspond more closely to the 3.75\,$\mu$T and 480\,$\mu$T cases for the Earth-like planet, respectively.

For K2-18b, the total SEP energy flux is $\sim$117 times larger than the total GCR energy flux. Although the host star K2-18 produces fewer SEPs than the Sun in our model, the SEP energy flux at the planet is significantly higher due to its close orbital distance. In the absence of a magnetic field, the GCR and SEP energy fluxes would become equal at a distance of $\sim$1.72\,au for a K2-18b-like planet. For magnetic field strengths of 1.7\,$\mu$T and 300\,$\mu$T, the GCR energy flux reduced by 11\% and 78\%, respectively, while the SEP energy flux decreases by 60\% and 95\%, shifting the crossover distance at which the energy fluxes would become equal to 1.16\,au and 0.84\,au. \citet{Rodgers-Lee+2023} find equal GCR and SEP molecular hydrogen ionization rates only at a distance of 10\,au for GJ~436 b. This is somewhat  surprising, given that the SEP flux adopted in this work for K2-18b is based on their model. Currently, it is unclear whether this discrepancy arises because the energy flux is not a reasonable proxy for the ionization rate, or from differences in the implementation of stellar wind effects, which may alter the resulting GCR and SEP spectra. Since the difference between 1.72\,au and 10\,au corresponds to a factor of $\sim$34 in flux, it is likely that both effects contribute to the discrepancy.

\begin{figure}
  \resizebox{\hsize}{!}{\includegraphics{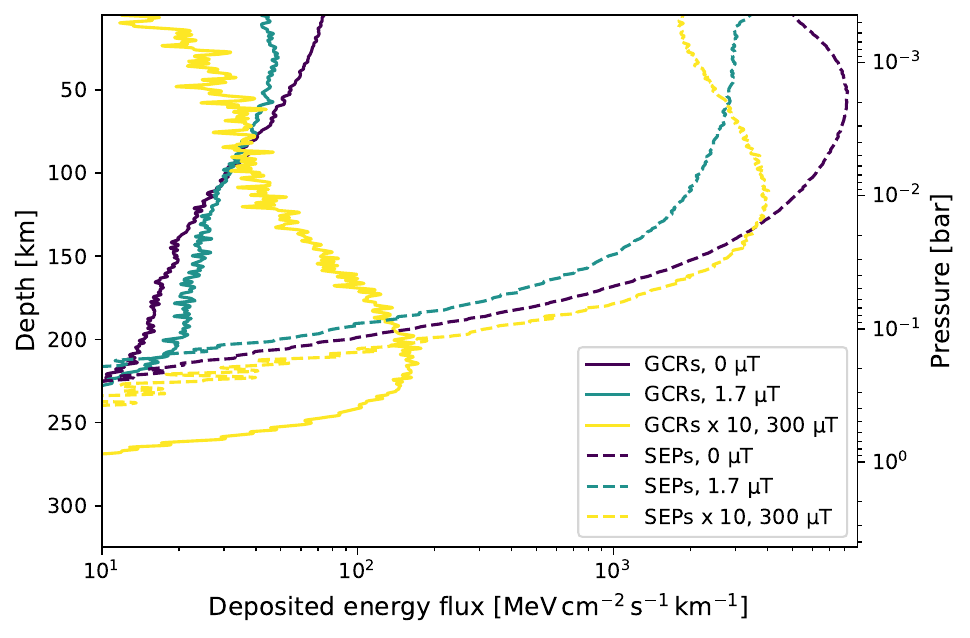}}
  \caption{Atmospheric energy deposition profiles of GCRs and SEPs in the atmosphere of K2-18b for different magnetic field strengths. The profiles have been smoothed to improve visual clarity. For visibility, the deposited energy corresponding to B=300\,$\mu$T has been multiplied by a factor of 10.}
  \label{Fig:K218bComparisonDeposition}
\end{figure}

The energy deposition for the different magnetic field strengths for K2-18b is shown in Fig. \ref{Fig:K218bComparisonDeposition}, again using  10$^6$\,particles for the GCR spectrum and 10$^7$\,particles for the SEP spectrum. The resulting energy deposition profiles differ from those shown in Fig. \ref{Fig:AtmosphereCompositionComparison} for an Earth-like planet. This is in part due to the larger atmospheric scale height of K2-18b, leading to a more gradual energy loss with altitude. In addition, the planetary magnetic field affects the energy deposition profile: in general, a larger fraction of the incident flux consists of higher-energy particles, which penetrate deeper into the atmosphere. This effect is particularly evident in the 300\,$\mu$T case for both GCRs and SEPs in Fig. \ref{Fig:K218bComparisonDeposition}.

We find that none of the simulated particles penetrate beyond $\sim$300\,km altitude, corresponding to roughly 2\,bar. This implies that, under the present conditions, the atmosphere is sufficiently thick to fully attenuate both GCRs and SEPs before they can reach the surface. However, this also suggests that particle radiation could,  in principle, reach the surface in cases where the atmosphere is as thin as $\sim$1\,bar \citep{Madhusudhan+2020,Piette+2020}. In contrast, \citet{Wogan+2024} argue that K2-18b is more likely to be a mini-Neptune with a substantially  thicker atmosphere. Although the atmospheric composition and scale height differ from those of Earth, they do not significantly change the pressure level at which GCR energy deposition becomes dominant over SEP energy deposition. This transition occurs at $\sim$10$^{-1}$\,bar, similar to the Earth-like case without a magnetic field. A strong dependence is observed for the magnetic field. Increasing magnetic field strength reduces the flux of incoming charged particles through magnetic deflection and rigidity cut-offs. This effect is energy dependent: low-energy SEPs are most efficiently excluded, while higher-energy particles are less affected and can still reach the atmosphere, often at higher latitudes, which was shown in Sect. \ref{Sect:Mag_field_dep}. Since SEPs typically dominate at lower energies than GCRs, the magnetic field suppresses SEPs more strongly than GCRs, shifting the crossover between the two populations to lower pressures (higher altitudes).

In addition, the magnetic field modifies the shape of the deposition profiles. By preferentially filtering out low-energy particles, which deposit their energy higher in the atmosphere, it reduces energy deposition at low pressures and shifts the peak deposition to deeper atmospheric layers (higher pressures). At the same time, the remaining particle population is biased toward higher energies, which have larger penetration depths, further enhancing energy deposition at larger depths. This effect is more pronounced for GCRs because their spectrum extends more strongly into the rigidity range affected by magnetic shielding, leading to a larger relative change in their transmitted population. Overall, the magnetic field therefore both reduces the total incoming flux and redistributes the surviving particle spectrum toward higher energies and higher latitudes, producing deeper and more concentrated energy deposition.

Our findings for K2-18b are broadly consistent with the results of \citet{Barth+2021} for HD 189733 b and \citet{Rodgers-Lee+2023} for GJ~436 b. In both studies, GCRs dominate over SEPs at higher pressures than found here, although this can largely be attributed to their use of electron production rates and hydrogen ionization rates, respectively, rather than energy fluxes, as well as differences in planetary parameters. Larger differences arise when considering the absolute pressure limits of cosmic-ray penetration. \citet{Barth+2021} find that GCRs penetrate the atmosphere beyond a pressure of 10$^1$\,bar, with SEPs reaching several bar, while \citet{Rodgers-Lee+2023} report penetration depths of $\sim$10$^3$\,bar for GCRs $\sim$10$^1$\,bar for SEPs. These values differ significantly from our results, where GCRs reach only $\sim$2\,bar and SEPs remain below 1\,bar.

Determining the maximum pressure limit to which cosmic rays can penetrate is inherently challenging, as the process is stochastic and therefore does not exhibit a sharp cut-off energy (depth). In addition, computational constraints limit the number of particles that can be simulated with full atmospheric complexity in {\tt Geant4}. \citet{Barth+2021} employ a fully parameterized ionization rate based on column density, whereas \citet{Rodgers-Lee+2023} use a custom Monte Carlo approach in which energy loss is parametrised as a function of particle energy. Such methods necessarily simplify the complex processes of atmospheric particle cascades and do not explicitly account for the production and transport of secondary particles. Furthermore, effects such as multiple scattering and the re-distribution of energy within particle showers might not be treated properly. Our results indicate that these processes are important, particularly at higher pressures, and that simplified approaches may therefore overestimate the atmospheric depth to which cosmic rays can penetrate. Given that GCRs and SEPs exhibit substantially different energy spectra, the resulting biases need not be the same for the two populations. For example, \citet{Rodgers-Lee+2023} find that GCRs penetrate to significantly higher pressures than SEPs, whereas we find a much smaller difference between the two.

\section{Discussion}
\subsection{Atmospheric chemistry}
In this work, we consider only the atmospheric energy deposition produced by cosmic rays, including the contribution from secondary particles generated in atmospheric cascades. This quantity provides a valuable first-order indication of the atmospheric layers affected by particle interactions and of the depths to which different cosmic-ray populations (GCR, SEP) can penetrate. However, the resulting chemical response of the atmosphere depends more directly on the ionization rate than on the deposited energy alone. Consequently, studying the impact of cosmic rays on atmospheric composition and on the interpretation of potential biosignatures requires an additional modelling step in which ionization rates are derived from the deposited energies. Such an approach was adopted by \citet{Rodgers-Lee+2023}, while the resulting chemical consequences have been investigated by, for example, \citet{Griessmeier+2016} and \citet{Barth+2021}. However, these studies relied on more simplified treatments of particle transport through the atmosphere than employed in the present work.

Our decision to restrict the present analysis to energy deposition was motivated by the desire to first establish a clear physical understanding of how cosmic-ray interactions depend on planetary and atmospheric properties. The interpretation of particle spectra and energy-deposition profiles is considerable more direct than that of the resulting chemical abundances, which are influenced by complex and often nonlinear chemical networks. This distinction is particularly important in the context of the present study, where we investigate the dependence of cosmic-ray interactions on magnetic-field strength, orbital distance, atmospheric pressure, and atmospheric composition. By focussing on energy deposition, we are able to isolate and quantify the influence of these parameters without introducing additional uncertainties associated with atmospheric chemistry. In future work, we will extend this framework by deriving ionization rates from the particle transport calculations and coupling them to chemical network models to assess the resulting effects on atmospheric composition and observable biosignatures.

For simplicity, we further assume that energy deposited at different latitudes produces an equivalent atmospheric response and therefore average over the planetary surface. This assumption is unlikely to hold in detail, particularly in the presence of a planetary magnetic field, which can introduce strong spatial variations in particle fluxes. A comprehensive assessment of the atmospheric and observational consequences of cosmic-ray irradiation therefore requires three-dimensional general circulation models coupled to atmospheric chemistry. For example, \citet{Chen+2021} studied the impact of stellar flares using such a model, albeit with a simplified treatment of magnetic shielding. An additional advantage of general circulation models is their ability to capture longitudinal asymmetries. As discussed in Sect.~\ref{Sec:Simplifications}, we assume that both GCRs and SEPs arrive isotropically at the planet. While this approximation is likely reasonable for GCRs, it is expected to be less accurate for SEPs, whose fluxes may exhibit substantial directional and temporal variability, particularly for close-in planets. Consequently, the atmospheric response may be strongly longitude dependent, an effect that cannot be captured within the framework adopted here.

\subsection{Surface interactions}
This study is focussed on the atmospheric energy deposition produced by cosmic rays, whereas studies of cosmic rays on Earth have often concentrated on the fraction of particles that reaches the planetary surface \citep{GossePhillips2001, BiermanEtAl2021, SchaeferEtAl2022}. The surface particle flux is of particular interest for studies of the potential impact of cosmic rays on surface habitability and biological systems. The results presented here can be used to estimate the fraction of incident cosmic-ray energy that reaches the surface as a function of (among others) atmospheric column density, atmospheric chemical composition, magnetic field, and orbital parameters. However, applying such estimates to exoplanets is challenging because atmospheric pressures and therefore atmospheric column densities are generally poorly constrained observationally. The same is true for exoplanetary magnetic fields. Since atmospheric shielding has a strong influence on the penetration depth of cosmic-rays, uncertainties in atmospheric structure directly translate into uncertainties in the surface radiation environment.

A comprehensive assessment of the effects of cosmic-rays on potential surface life also requires knowledge of the particle composition and energy spectrum of the incident radiation, as biological effects depend not only on the total energy flux but also on the types and properties of the individual particles and the secondary cascades they generate. In the present study, we focus primarily on the total energy flux and its deposition within the atmosphere. Future work should therefore combine detailed particle transport calculations with realistic atmospheric models and biologically relevant radiation metrics to evaluate the impact of cosmic rays on the habitability of specific exoplanets.

\section{Summary} \label{Sec:Summary}
In this study, we modelled the interaction of cosmic rays with planetary atmospheres as a function of planetary magnetic field strength, orbital distance from the host star, atmospheric chemical composition, and atmospheric pressure. We also investigated the influence of the solar modulation potential on cosmic-ray energy deposition. Specifically, the model was applied to both an Earth-like reference planet and the exoplanet K2-18b as a representative case study. The studies were performed using the state-of-the-art nuclear reaction toolkit {\tt Geant4}.

Our analysis is focussed on the atmospheric energy deposition produced by cosmic rays, including the contribution from secondary particles generated in atmospheric cascades. This quantity provides a useful first-order measure of the atmospheric layers affected by particle interactions and of the depths to which different cosmic-ray populations (GCR, SEP) can penetrate. Although energy deposition alone does not capture the full complexity of cosmic-ray-induced atmospheric chemistry and dynamics, it constitutes a fundamental step towards understanding the influence of high-energy particles on planetary atmospheres. The main results or our study can be summarized as follows:
\begin{itemize}
    \item GCRs and SEPs respond differently to planetary magnetic fields, because the GCR energy flux is dominated by particles with energies above 1\,GeV, whereas SEP particles are predominantly below this threshold. Consequently, magnetic fields enhance the GCR energy flux at high latitudes, while no such enhancement is found for SEPs. At lower latitudes, magnetic fields reduce the energy flux in both cases, with a substantially stronger effect for SEPs.
    \item In the absence of a magnetic field, the globally averaged energy flux of GCRs and SEPs becomes equal at an orbital distance of $\sim$2.4\,au. For a planetary magnetic field strength of 30\,$\mu$T, this distance decreases to approximately 1\,au. Stronger magnetic fields produce little additional change, indicating a saturation of the shielding effect.
    \item The stellar modulation parameter influences the GCR energy flux, although the effect remains limited because it primarily affects lower-energy particles. For both GCRs and SEPs, atmospheric composition has only a minor impact on the pressure levels at which cosmic-ray energy is deposited.
    \item For K2-18b, the influence of the magnetic field is slightly stronger than for the Earth-like reference case, likely due to the planet's larger radius. Cosmic rays penetrate to pressure levels comparable to those found for the Earth-like atmosphere. These pressures are significantly lower than those reported in previous studies employing simplified treatments of atmospheric cosmic-ray interactions.
    \item Our results demonstrate that planetary magnetic fields strongly modulate the spatial distribution of cosmic-ray energy deposition, particularly for SEPs. Accurate assessments of cosmic-ray-driven atmospheric chemistry and climate effects therefore require realistic magnetic-field models in addition to detailed descriptions of atmospheric particle interactions.
\end{itemize}

Overall, we find that planetary magnetic fields play a key role in shaping the cosmic-ray environment of planetary atmospheres. While magnetic shielding has a limited influence on the globally averaged GCR energy flux, it strongly affects the latitudinal distribution of both GCR and SEP and, therefore, the spatial location of energy deposition within the atmosphere. These results highlight the importance of, first, using sophisticated and well-tested particle transport and nuclear interaction models, and second, incorporating realistic magnetic-field geometries into such models for studies of impact of cosmic rays on atmospheric chemistry, climate, and habitability.

Future works should focus on improving our understanding of exoplanetary magnetic fields and on coupling advanced cosmic-ray transport models to three-dimensional atmospheric circulation and chemistry models. Such developments will be essential for accurately quantifying the impact of cosmic rays on exoplanet atmospheres and their observational signatures. Efforts in these directions are already underway and are expected to provide a more comprehensive framework for assessing the role of cosmic rays in shaping planetary atmospheres.

\begin{acknowledgements}
We thank the anonymous referee for useful comments and suggested additions to this paper. J.P acknowledges the support from the Swiss National Science Foundation under grant 200021\_204847.
\end{acknowledgements}

%
%

\bibliographystyle{aa}
\bibliography{bibliography.bib}

\begin{appendix} 
\nolinenumbers
\section{Galactic and stellar cosmic ray spectra} \label{Sect:spectra}
The SEP and GCR particle spectra constitute the primary input parameters of the model. We first describe the GCR particle spectra adopted in this work. \citet{Leyaetal_2021} discussed the different local interstellar spectra (LIS) and the heliospheric transport and modulation process. For primary Galactic protons, we use the spectrum originally developed by \citet{Castagnoli1980}, and later revised by \citet{MasarikReedy_1996}. For more information we refer to \citet{Hirtz2019}. This calculation is expressed as

\begin{equation} 
    J_{p}(T,M)=c_p \frac{T (T+2m_{p}c^2) (T+x+M)^{-2.65}}{(T+M) (T+2 m_{p}c^{2}+M)},
\end{equation}

\noindent with $x$ defined as

\begin{equation}
    x = 780 e^{-2.5 \cdot 10^{-4} \cdot  T}.
\end{equation}

\noindent Here, $T$ is the kinetic energy, $M$ the solar modulation parameter, and m$_p$ the proton rest mass. Using a normalization factor c$_p$=10$^5$ yields differential fluxes in units of cm$^{-2}\,$s$^{-1}\,$sr$^{-1}\,$MeV$^{-1}$. For primary Galactic $\alpha$ particles, \citet{Hirtz2019} developed a parameterized spectrum based on the assumption that the spectrum for primary Galactic $\alpha$-particles is very similar to the spectrum for primary Galactic protons, if both are discussed in terms of energy per nucleon. Galactic cosmic rays consist predominately of protons ($\sim$87\%) and $\alpha$ particles ($\sim$12\%). This is expressed as

\begin{multline}
       J_{\alpha}(T,K) = c_{\alpha}T^{K}(T+2 m_{\alpha}c^{2})/\\((T+700)(T+2m_{\alpha}+700)(T+312500T^{-2.5}+700)^{1.65+K}),
\end{multline}

\noindent where the parameter $K$ is related to the solar modulation potential $M$ through

\begin{equation}
    K=(1.786\cdot10^{-3}M)-0.1323.
\end{equation}  

\noindent Here, $T$ denotes the kinetic energy per nucleon, m$_\alpha$ the alpha particle mass, and c$_\alpha$=5.5$\cdot$10$^3$ yields fluxes in units of cm$^{-2}\,$s$^{-1}\,$sr$^{-1}\,$MeV$^{-1}\,$nucleon. This parametrization was introduced primarily for computational convenience, as it allows both proton and $\alpha$-particle spectra to be specified through one single value for the solar modulation potential, $M$. It should therefore be regarded as a practical model input rather than as the direct result of dedicated heliospheric transport and modulation calculations.

For the stellar spectrum we consider only protons. The differential flux is parameterized as

\begin{equation}
    J(R)=A_0 e^{-R/R_0},
\end{equation}

\noindent following \citet{McGuire+1984}. Here, $R_0$=90\,MV denotes the characteristic rigidity, and $A_0$=61.1 corresponds to the normalization derived from the average solar proton spectrum at a heliocentric distance of 1 au. Both the characteristic rigidity and the absolute proton flux, $J$, are not well constrained. For instance, \citet{Nishiizumietal_2009} report values of $R_0 = 90$\,MV and $J = 73$ protons$\,$cm$^{-2}\,$s$^{-1}\,$(4$\pi$)$^{-1}$ for energies above 10\,MeV, based on their analysis of the solar cosmic ray record in lunar rock 64455.

\section{Particle propagation and neutron production}
\begin{figure}
  \resizebox{\hsize}{!}{\includegraphics{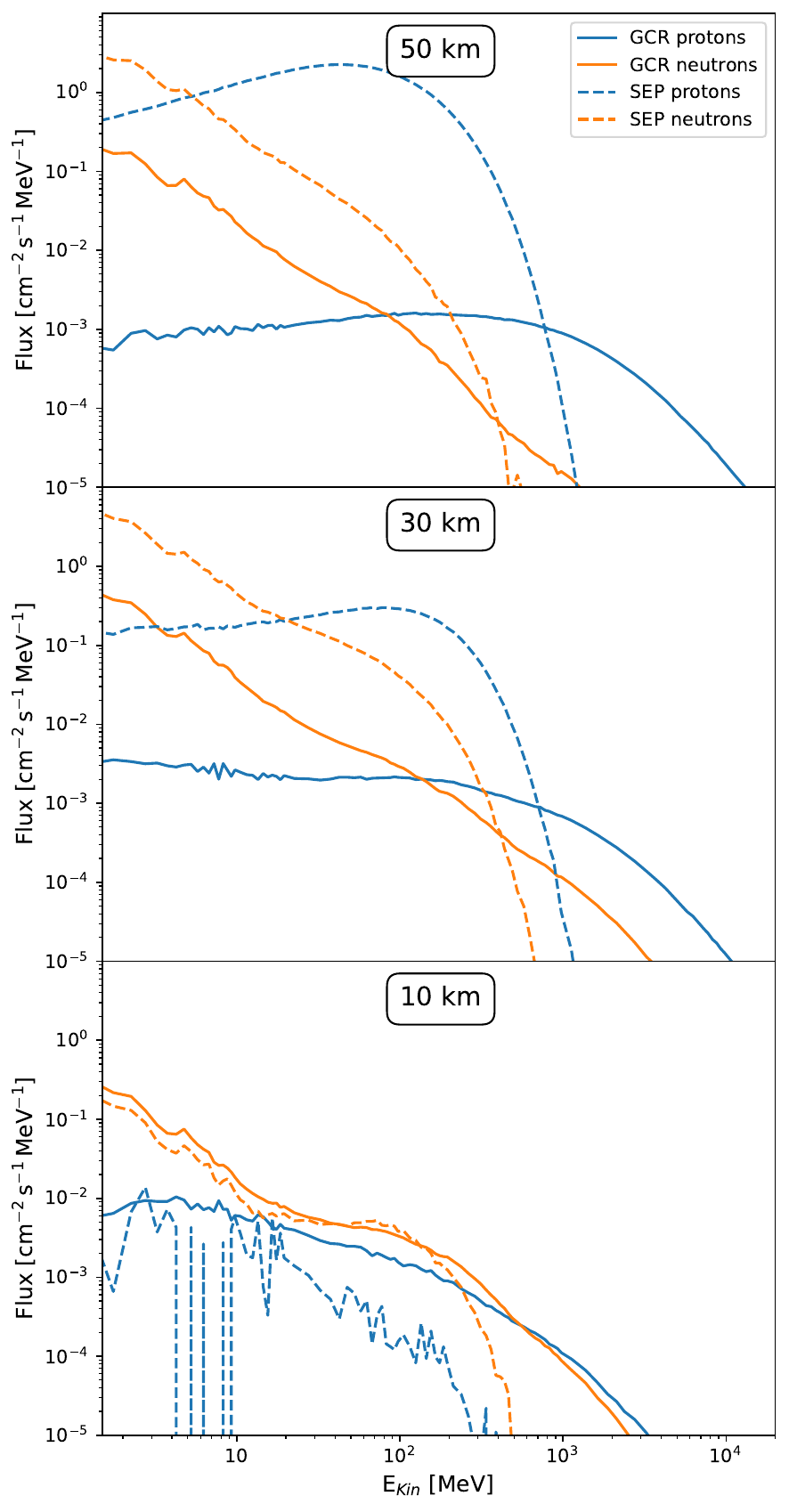}}
  \caption{Proton and neutron fluxes of GCRs and SEPs at altitudes of 50\,km (top panel), 30\,km (middle panel), and 10\,km (bottom panel) in the atmosphere of an Earth-like planet without magnetic field at a distance from the host star of 1\,au.}
  \label{Fig:SpectrumAltitudeComparison}
\end{figure}

Figure \ref{Fig:SpectrumAltitudeComparison} depicts the GCR and SEP proton and neutron fluxes at different altitudes in the atmosphere. The primary GCR spectrum consists of protons and alpha-particles, while all neutrons are secondary particles produced through interactions of primary cosmic rays with the atmosphere. For both GCRs and SEPs, the neutron flux increases between 50 and 30\,km altitude as a result of proton-induced nuclear reactions in the atmosphere. For SEPs, the proton flux decreases at all energies, whereas for GCRs the low-energy proton flux increases slightly due to the production of secondary protons by higher-energy particles. In general, secondary-particle production is much more pronounced for GCRs than for SEPs because of the higher energies of the primary GCRs particles. Between 30 and\,10 km altitude, both the proton and neutron flux decrease for GCRs and SEPs. However, the decrease in neutron flux is substantially smaller than that of the proton flux because  neutrons, being electrically neutral, do not undergo continuous energy losses through electronic interactions and therefore penetrate further into the atmosphere. 

\begin{figure}
  \resizebox{\hsize}{!}{\includegraphics{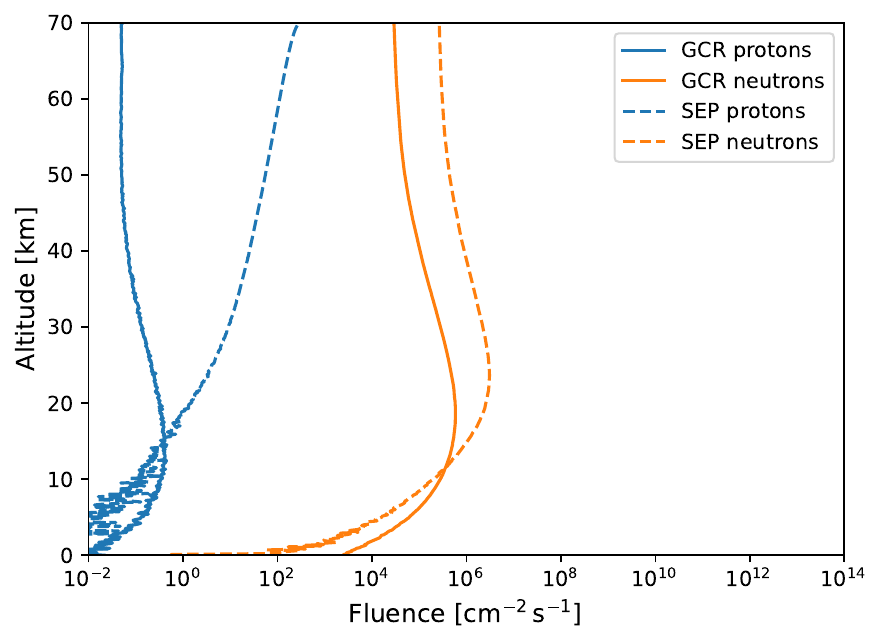}}
  \caption{Neutron and proton particle fluence of GCRs and SEPs as a function of altitude. Only particles with energies above  1\,MeV are considered for calculating the fluences.}
  \label{Fig:ParticleFluenceComparison}
\end{figure}

Figure \ref{Fig:ParticleFluenceComparison} shows the GCR and SEP proton and neutron fluence throughout the atmosphere, restricted to particles with energies above 1\,MeV. The continuous decrease in the SEP proton fluence and the increase in GCR proton fluence are consistent with the results shown in Fig. \ref{Fig:SpectrumAltitudeComparison}. As mentioned above, the difference is due to the different efficiencies in producing secondary particles. The figure also clearly shows an increase in the neutron fluence for both GCRs and SEPs, as well as a slower decrease with altitude compared to protons, reflecting the more efficient atmospheric penetration of neutrons.

A lower energy threshold of 1\,MeV was adopted to mitigate uncertainties in the lowest-energy part of the fluence spectrum while ensuring a consistent comparison between proton and neutron populations. We find that lowering this threshold to, for example, 1\,keV does not affect the altitude dependence of neutron fluences for either particle source. However, it increases the absolute neutron fluence by more than four orders of magnitude, clearly demonstrating the strong contribution of low-energy secondary particles to the total neutron production. 1\,keV is well above typical atomic ionization energies.

\end{appendix}
\end{document}